\documentclass[a4paper,11pt]{article}
\pdfoutput=1 

\usepackage{jheppub} 

\usepackage[T1]{fontenc} 

\usepackage{hyperref}
\usepackage{tensor,mathrsfs}

\title{\boldmath JT gravity from holographic reduction of 3D asymptotically flat spacetime}

\author[1,2]{Arindam Bhattacharjee,}
\author[3]{Muktajyoti Saha}

\affiliation[1]{Harish-Chandra Research Institute,\\ Chhatnag Road, Jhunsi, Prayagraj—211019, India}
\affiliation[2]{Homi Bhabha National Institute,\\ Training School Complex, Anushaktinagar, Mumbai 400094, India}
\affiliation[3]{Indian Institute of Science Education and Research Bhopal,\\ Bhopal Bypass, Bhauri, Bhopal 462066, India}

\emailAdd{arindamb.hep@gmail.com} 
\emailAdd{muktajyoti17@iiserb.ac.in}

\abstract{We attempt to understand the CFT$_1$ structure underlying (2+1)D gravity in flat spacetime via dimensional reduction. We observe that under superrotation, the hyperbolic (and dS$_2$) slices of flat spacetime transform to asymptotically (A)dS$_2$ slices. We consider a wedge region bounded by two such surfaces as End-of-the-World branes and employ Wedge holography to perform holographic reduction. We show that once we consider fluctuating branes, the localised theory on the branes is Jackiw-Teitelboim (JT) theory. Finally, using the dual description of JT, we derive an 1D Schwarzian theory at the spatial slice of null infinity. In this dual Celestial (nearly) CFT, the superrotation mode of 3D plays the role of the Schwarzian derivative of the boundary time reparametrization mode.}

\begin{document} 
\maketitle
\flushbottom
\section{Introduction}
The nature of holography in flat spacetime has been a subject of intense study in recent years (See \cite{Strominger:2017zoo,Pasterski:2016qvg,Laddha:2020kvp} and references therein). Analysing the structure of asymptotic symmetries in 4D flat spacetimes \cite{Kapec:2016jld, Barnich:2009se}, it has been argued \cite{Pasterski:2016qvg} that the dual theory at the boundary is a two dimensional CFT, termed Celestial CFT. More generally, it was argued that (d+1) dimensional theory of gravity in asymptotically flat spacetime has a (d-1) dimensional CFT description \cite{Pasterski:2017kqt}.
\vspace{10pt}\\
At first glance, this seems pretty straightforward. Since the Lorentz group of (d+1) dimensional flat spacetime is $SO(d,1)$ which is also the global conformal group at (d-1) dimensions. Thus every flat space amplitude, by virtue of being Lorentz invariant, is also conformally invariant. This invariance can be made explicit when the amplitudes are written in appropriate basis. The non-triviality comes while enhancing the global conformal invariance of the 4D amplitudes to full Virasoro invariance of the dual 2D CFT description \cite{Barnich:2010ojg}. These infinite dimensional symmetries are interpreted in the bulk as the superrotation symmetries associated with subleading soft graviton theorem \cite{Campiglia:2014yka, Kapec:2014opa}. Thus the amplitudes of any quantum field theory which is coupled to gravity in the bulk are constrained by Ward identities coming from the full Virasoro symmetry. This gives an extraordinary advantage over the structure of generic gauge/gravity amplitudes and produce new insights into their properties.
\vspace{10pt}\\
A similar story should follow for 3D gravity in asymptotically flat spacetimes. The dual theory is now expected to be a 1-dimensional CFT with one copy of Virasoro algebra as its symmetry. In the bulk description, the Virasoro symmetry is a part of the asymptotic symmetry group BMS$_3$. But gravity in (2+1) dimensions is a topological theory with no gravitons propagating in the bulk. Thus the interpretation of the asymptotic symmetries in terms of soft graviton theorems are lost and a special care is needed to understand the BMS$_3$/CFT$_1$ correspondence \cite{Barnich:2010eb}. In this paper we approach this problem via foliating (2+1)D flat spacetimes into hyperbolic (and dS$_2$) slices and then reducing the theory in these slices. In (3+1)D, a similar approach was followed by \cite{deBoer:2003vf, Cheung:2016iub}. But as we show in the paper, the topological nature of gravity in (2+1)D forces us to consider the boundary behaviour of the theories much more carefully.
\vspace{10pt}\\
As we see below, the hyperbolic slicing foliates the (Milne part of) (2+1)D flat spacetime into warped product of AdS$_2$ and $\mathbb{R}$. Since gravity is not Weyl invariant, the reduction to a particular AdS$_2$ slice becomes involved. To do so explicitly, we invoke the recently put forward idea of Wedge Holography \cite{akal_codimension_2020}. In this setup, we study the bulk gravity theory in a wedge region bounded by two hyperbolic slices (called the "End-of-the-world branes") and reduce it to its boundary. We show that superrotation symmetries transform the boundaries of this wedge region from Euclidean AdS$_2$(EAdS) spaces to asymptotically EAdS spaces.
\vspace{10pt}\\
To study localised dynamics on these End-of-the-world (EoW) branes we turn on fluctuations. First, to respect superrotation symmetry, we only allow fluctuations along the spatial slices keeping their position rigidly fixed. Considering only the massless sector, these fluctuations correspond to pure gravity in 2D which is known to be non-dynamical. Finally, we break the superrotation symmetry by introducing fluctuating branes. The scalar mode associated with this fluctuation couples non-minimally with the 2D gravity action above and hence behaves as a dilaton. The complete theory on the brane turns out to be JT gravity \cite{Jackiw:1984je, Teitelboim:1983ux}. The technique of conformal symmetry breaking through brane fluctuations was introduced by \cite{geng_jackiw-teitelboim_2022, geng_aspects_2022} in AdS$_3$ context. In these works, the effective theory on the branes was also JT gravity (See also \cite{Deng:2022yll}, for similar results in AdS$_3$). Although there are no local bulk degrees of freedom in this effective theory, there are nontrivial boundary modes, which become extremely important in the context of holography \cite{Maldacena:1998uz,maldacena_conformal_2016}.
\vspace{10pt}\\
One of the most interesting results we arrive at through this procedure of holographic reduction is to identify the Schwarzian action dual to JT gravity as the effective action for superrotation modes.\footnote{This is similar to the Schwarzian part of the action in \cite{carlip_dynamics_2017}.} In 4D, the 2D CFT whose stress tensor are the superrotation modes is dubbed Celestial CFT. Following that, the Schwarzian action can be interpreted as the Celestial (nearly) CFT dual to (2+1)D gravity in asymptotically flat spaces. The effective JT gravity description on the boundary of the wedge region makes this duality manifest. We also show that the superrotation mode in 3D plays the role of the Schwarzian derivative of the 2D boundary graviton mode. This is a crucial result and a concrete realisation of BMS$_3$/CFT$_1$ correspondence \cite{Barnich:2010eb}.
\vspace{10pt}\\
The content of the paper is organised as follows: In section \ref{asymp-sym}, we discuss how superrotation takes (A)dS$_2$ slices to asymptotically (A)dS$_2$ hypersurfaces in 3D. In section \ref{reduction}, we dimensionally reduce the 3D pure gravity theory on (A)dS$_2$ slices and obtain JT gravity in the low energy limit. In section \ref{bdy-theory}, we find an effective Schwarzian theory that lives on any spatial slice of the future null infinity and identify the 3D superrotation mode as the Schwarzian. Finally in section \ref{concl}, we summarize the results of the paper.

\section{Foliations and their relation to asymptotic symmetry} \label{asymp-sym}
\subsection{Foliations of flat spacetime}
\begin{figure}[h]
\centering
\includegraphics[width=0.5\textwidth]{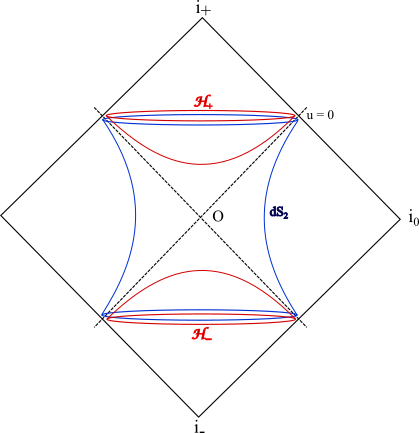}
\caption{Penrose diagram of flat space and (A)dS$_2$ foliations. The $\mathcal{H}_{\pm}$ slices are at constant timelike separation from the origin O. The dS$_2$ slices are at constant spacelike separation.}
\label{foliation}
\end{figure}

In this section, we start by foliating the flat space into hyperbolic (and dS$_2$) slices. The basic idea is to fix an origin in the bulk and consider hypersurfaces that are at fixed timelike or spacelike separations from this origin.  This divides the full spacetime into three regions, covered by different coordinate patches, as we see below.\\
The metric in (2+1) dimensional flat spacetime $\mathbb{R}^{2,1}$ is given by,
\begin{align}
    ds^2 = -dt^2 + dr^2 + r^2d\theta^2, \label{trflat}
\end{align}
with $-\infty<t<\infty$, $0<r<\infty$, and $0\leq\theta\leq 2\pi$. Now fixing an origin O (See Figure \ref{foliation}) we identify the following three regions:
\begin{itemize}
    \item The region inside the future light cone (denoted as $\mathcal{H}^+$) i.e. $0<t<\infty$ can be realized as a foliation of two-dimensional hyperbolic slices. We perform the following coordinate transformations,
    \begin{align}
        & t = \tau\cosh{\rho}, \quad r = \tau\sinh{\rho}; \qquad 0<\tau<\infty, \quad 0<\rho<\infty, \\
        & ds^2 = -d\tau^2 + \tau^2(d\rho^2 + \sinh^2{\rho}d\theta^2). \label{hyper}
    \end{align}
    The constant $\tau$ slices are Euclidean $AdS_2$.
    \item The region inside the past light cone (denoted as $\mathcal{H}^-$) i.e. $-\infty<t<0$ has the same structure as $\mathcal{H}^+$.
    \item The region outside the light cone can be realized as a foliation of two-dimensional de Sitter slices. We perform the following coordinate transformations,
    \begin{align}
        & t = \xi\sinh{\eta}, \quad r = \xi\cosh{\eta}; \qquad 0<\xi<\infty, \quad -\infty<\eta<\infty, \\
        & ds^2 = d\xi^2 + \xi^2(-d\eta^2 + \cosh^2{\eta}d\theta^2). \label{dS}
    \end{align}
    The constant $\xi$ slices are Lorentzian $dS_2$. The metric \eqref{dS} in de Sitter slicing can be obtained from the metric \eqref{hyper} in hyperbolic patch, through the following analytic continuation of coordinates:
    \begin{align}
        \tau = -i\xi, \quad \rho = \eta - i\frac{\pi}{2}. \label{hyper2dS}
    \end{align}
\end{itemize}
The region inside the light cone is called the Milne patch and the region outside is called the Rindler patch.
\subsection{(2+1)D asymptotically flat spacetimes}
Asymptotically flat spacetimes were first studied in 4-dimensions and it was realised that the symmetry group near null infinity is enhanced from Poincar\`{e} to an infinite dimensional BMS$_4$ group\cite{Bondi:1962px,Sachs:1962wk}. This consists of a semi-direct product of the Lorentz group with an infinite dimensional abelian generalisation of translations (known as "super-translations"). It was later enhanced to include local conformal transformations on the asymptotic 2-sphere instead of just Lorentz group\footnote{Further enhancement to diffeomorphisms on $S^2$ have been considered in \cite{Campiglia:2014yka, Campiglia:2015yka}.} \cite{Barnich:2010ojg}. These local conformal transformations were termed "superrotations". The whole enhanced BMS$_4$ group is conjectured to be a symmetry of the quantum gravitational S-Matrix in 4D \cite{Strominger:2013jfa, Kapec:2014opa}.
\vspace{10pt}\\
The story in 3D follows a similar pattern \cite{Barnich:2010eb}. As we will extensively use the asymptotically flat 3D metric, here we briefly sketch its form.
\vspace{10pt}\\
In Bondi gauge, any metric can be written in the form:
\begin{align}\label{bms-gauge}
    ds^2 = e^{2\beta} \frac{V}{r} du^2 - 2 e^{2\beta} du\,dr + r^2 (d\theta -Udu)^2,
\end{align}
where the parameters $\beta,V,U$ are functions of $(u,r,\theta)$.
In this gauge the flat space is given by $\frac{V}{r} = -1; \quad \beta = 0; \quad U =0$.\\
Demanding asymptotic flatness, we reach the following form of the metric\footnote{In principle the S$^1$ part of the metric written below can also have a non-trivial scaling factor, but we consider only meromorphic functions on S$^1$ for which this factor can be absorbed into the redefinition of the angular co-ordinate.}:
\begin{align}
    ds^2 = -\mathcal{M}(u,\theta)du^2 - 2dudr -  \mathcal{N}(u,\theta) dud\theta + r^2 d\theta^2, \label{3D_Bondi_almost}
\end{align}
and Einstein's equations further dictate $\partial_u \mathcal{M} = 0; \quad \partial_{\theta} \mathcal{M} = \partial_u \mathcal{N}$. Thus a general solution can be written as:
\begin{align}
    ds^2 = -M(\theta)du^2 - 2dudr - (uM'(\theta)+J(\theta))dud\theta + r^2 d\theta^2. \label{3D_Bondi}
\end{align}
The functions $\{M(\theta), J(\theta)\}$ span the phase space of solutions of 3D asymptotically flat metric. At this point, we should compare this metric with its 4D variant. The $u\theta$ component of the metric contains non-trivial data at $r^{(0)}$ order in 3D. This is not the case in 4D. There, the corresponding metric component ($g_{uz}, g_{u\bar{z}}$) is completely determined at $r^{(0)}$ order and non-trivial data comes at order $1/r$.\\
The asymptotic symmetry algebra for this class of metrics is called the BMS$_3$ algebra. It is generated by the asymptotic Killing vector $\chi^{A}$ that keeps the form of the metric \eqref{3D_Bondi} invariant up to leading order. It turns out that the form of the asymptotic Killing vector is \cite{barnich_dual_2013},
\begin{align}\label{killing-bms3}
\chi^u &= T(\theta) + u Y'(\theta), \nonumber\\
\chi^{\theta} &= Y(\theta) - \frac{1}{r} \partial_\theta\chi^u, \nonumber\\
\chi^r &= -r \partial_{\theta}\, \chi^{\theta} + \frac{1}{2r}(uM'(\theta)+J(\theta)) \partial_{\theta}\chi^u , 
\end{align}
with $T(\theta)$ and $Y(\theta)$ being arbitrary functions on $S^1$. The modes of $T(\theta)$ commute with each other and they are called \textit{supertranslations}. On the other hand, the modes of $Y(\theta)$ satisfy Witt algebra and are called \textit{superrotations}.
\vspace{10pt}\\
It is instructive to see how the asymptotically flat metric transforms under the action of $\chi_A$. Using $\delta G_{AB} = \nabla_A \chi_B + \nabla_B \chi_A$ we find,
\begin{align}
    \delta M &= Y(\theta) M'(\theta) + 2 M(\theta) Y'(\theta) + 2 Y'''(\theta), \label{var-of-M}\\
    \delta J &= T(\theta) M'(\theta) + 2 M(\theta) T'(\theta)+ 2 J(\theta) Y'(\theta) +Y(\theta) J'(\theta) + 2 T'''(\theta). \label{var-of-J}
\end{align}
These equations lend us remarkable insight. Firstly looking at the equation \eqref{var-of-M} we can identify the RHS as the infinitesimal Schwarzian derivative. Thus the function $M(\theta)$ behaves like a \textit{Schwarzian}. We will see important implications of this in our results.
Also notice from \eqref{var-of-J} that under superrotations, the $J(\theta) = 0$ spacetimes form a closed group. 
\subsection{Effect of superrotation on foliations}
We now consider purely superrotated spacetimes for which $J(\theta) = 0$. Thus the metric is specified by,
\begin{align}
    ds^2 = -M(\theta)du^2 - 2dudr - uM'(\theta)dud\theta + r^2 d\theta^2. \label{3D_superrotated}
\end{align}
The pure superrotated metrics can be written in a hyperbolic coordinate system using the following transformations,
\begin{align}
    u = \frac{\tau}{\sqrt{M(\theta)}}\text{e}^{-\rho}, \quad r = \tau\sqrt{M(\theta)}\sinh{\rho}.
\end{align}
The metric transforms as,
\begin{align}
    ds^2 = -d\tau^2 + \tau^2\left[d\rho^2 + \frac{M'}{M}d\rho d\theta + \left(M \sinh^2{\rho} + \frac{(M')^2}{4M^2}\right)d\theta^2\right]. 
\end{align}
We diagonalize the metric on $(\rho,\theta)$ surface by simply transforming $\rho = \tilde{\rho} - \ln{\sqrt{M}}$ such that the metric becomes,
\begin{align}
    ds^2 = -d\tau^2 + \tau^2\left[d\tilde{\rho}^2 + \left(\frac{1}{4}\text{e}^{2\tilde{\rho}} -\frac{M}{2} + \frac{M^2}{4}\text{e}^{-2\tilde{\rho}} \right) d\theta^2\right]. \label{Bondi_in_hyper}
\end{align}
The constant $\tau$ slices have the structure of asymptotically $AdS_2$ metric \cite{grumiller_menagerie_2017}. In terms of $(\tau,\tilde{\rho},\theta)$ coordinates, we have:
\begin{align}
    u = \tau\text{e}^{-\tilde{\rho}}, \quad r = \frac{\tau}{2}\text{e}^{\tilde{\rho}} - \frac{\tau M}{2}\text{e}^{-\tilde{\rho}}.
\end{align}
Thus we see that under superrotations, the Euclidean AdS$_2$ foliations transform to asymptotically AdS$_2$ slices. A similar transformation occurs when we consider dS$_2$ slices in the Rindler region of the 3D spacetime.\\
This is an important result and it shows that to understand holographic reduction of pure gravity in asymptotically flat spacetimes, it only makes sense to reduce the theory to asymptotically AdS$_2$(and dS$_2$) slices. Below we undertake that task.

\section{Dimensional Reduction of Einstein-Hilbert action} \label{reduction}
We start with Einstein-Hilbert action in 3D with zero cosmological constant,
\begin{align}
    S = \frac{1}{16\pi G_N}\int d^3\hat{x} \sqrt{-\hat{g}}\hat{R}, \label{Einsbulk}
\end{align}
which has flat space as a solution. The flat spacetime in the hyperbolic patch has the structure of a warped product of $\mathbb{R}$ and $AdS_2$ i.e. the radii of the $AdS_2$ surfaces vary with the coordinate $\tau$ on $\mathbb{R}$. We want to reduce the theory \eqref{Einsbulk} on an $AdS_2$ slice and then use the tools of holography in 2D, which would lead to a one-dimensional dual description.
\vspace{10pt}\\
In the standard Kaluza-Klein reduction, the background has a product structure, for instance, $\mathcal{M}\times\mathcal{N}$. Here $\mathcal{N}$ is some compact manifold. The key idea is to expand all the fields in some basis on $\mathcal{N}$ and integrate out the action over $\mathcal{N}$. This leads to a lower dimensional theory on $\mathcal{M}$, which typically has an infinite number of fields having some discrete labels. The lowest mode is massless, whereas all others have mass inversely proportional to the volume of the compact space. The massive modes can be integrated out to arrive at a low energy effective theory involving massless modes only.
\vspace{10pt}\\
But in our case, we want to reduce along the $\mathbb{R}_{\tau}$ direction which is non-compact. Also the background does not have a product structure. To circumvent similar issues, the idea of Wedge holography was used in $AdS_3$ \cite{geng_jackiw-teitelboim_2022,geng_aspects_2022}. Below we briefly review Wedge holography and then apply it to asymptotically flat spacetime.
\subsection{Wedge Holography: A brief review}
Wedge holography \cite{akal_codimension_2020,miao_exact_2020} is a realisation of co-dimension two holography where the dual theory of a (d+1) dimensional gravity theory lives in some (d-1) dimensional surface rather than the usual d-dimensional one. It was proposed as a generalisation of AdS/CFT and then later was used\footnote{Wedge holography has also been used in \cite{Geng:2020fxl, Geng:2021iyq} in understanding concepts related to black hole information paradox.} in understanding Celestial holography in flat spacetime\cite{ogawa_wedge_2022}. 
\vspace{10pt}\\
The basic idea is to consider a "wedge" region $W$ in a (d+1) dimensional bulk spacetime $\mathcal{M}$ bounded by two EoW branes $Q_1$ and $Q_2$. The part of the boundary $\partial\mathcal{M}$ of the full manifold within $W$, is denoted as $\Sigma$. The classical gravity in the wedge region is defined by specifying Neumann boundary conditions on $Q_1$ and $Q_2$ while the usual Dirichlet boundary conditions are imposed on $\Sigma$. Then the proposal of Wedge Holography states \cite{akal_codimension_2020}:
\vspace{5pt}\\
\textit{Classical gravity on the Wedge region is dual to a CFT living on $\Sigma$ in the limit when the width of $\Sigma$ is going to zero.}
\vspace{5pt}\\
In the above limit the surface $\Sigma$ becomes a (d-1) dimensional surface in which the CFT lives. Thus we have a co-dimension two holography.
\begin{figure}[h] \label{wedge-review}
\centering
\includegraphics[scale=0.25]{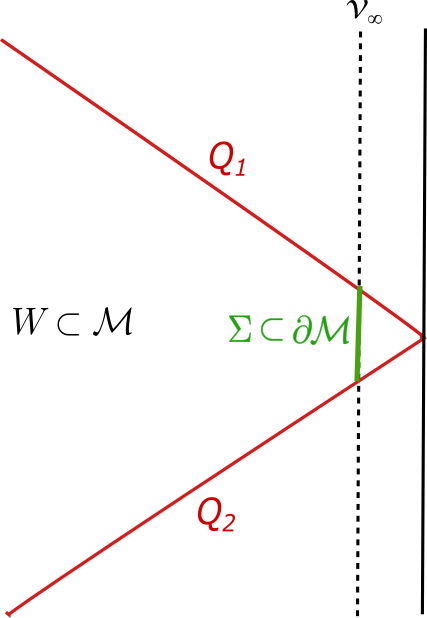}
\caption{The wedge holography setup. Here $\Sigma \xrightarrow{} 0$ is implemented through introducing a large cutoff surface $v_{\infty}$ near the boundary and then taking $v_{\infty} \xrightarrow{} \infty$.}
\end{figure}
In case of flat space, the wedge region is selected in the Milne region and in the Rindler region separately \cite{ogawa_wedge_2022}. In Milne patch, the wedge region is bounded by two AdS$_2$ slices with proper distance $\tau_1$ and $\tau_2$ from the origin O. Similarly, in Rindler patch, the wedge region is bounded by dS$_2$ slices at proper distance $\xi_1$ and $\xi_2$. In both cases, the (d-1) surface is then the $u=0$ surface at null infinity.\\
We employ wedge holography in these two regions separately and then match the boundary conditions on $u=0$. First, we keep the boundaries of the wedge region to be fixed and then allow fluctuating boundaries. In later case we get an effective JT theory on the EoW branes.

\subsection{Reduction to rigid slices}
As stated above, we consider a region $W$ within the Milne patch bounded by two $AdS_2$ slices at $\tau=\tau_1$ and $\tau=\tau_2$. The 3D gravity action supplemented by appropriate boundary terms with Neumann boundary condition is as follows,
\begin{align}
    S_W = \frac{1}{16\pi G_N}\Big[\int_W d^3\hat{x} \sqrt{-\hat{g}}\hat{R} & + 2\int_{Q_1}  d^2\hat{x}\sqrt{\hat{h}^{(1)}}(\hat{K}^{(1)} - T^{(1)}) \nonumber \\
    &- 2\int_{Q_2} d^2\hat{x}\sqrt{\hat{h}^{(2)}}(\hat{K}^{(2)} - T^{(2)})\Big]. \label{actW}
\end{align}
Here $Q_i$ is the boundary at $\tau = \tau_i$, with induced metric $\hat{h}^{(i)}$, outgoing unit normal $\hat{n}^{(i)}_A$, trace of extrinsic curvature $\hat{K}^{(i)}$, and "tension" $T^{(i)}$. \footnote{There is no boundary term for Neumann boundary condition, that makes the variation of the action zero on-shell. Hence, the "tension" is added to the boundary which corresponds to having some localized matter on the boundary. Here it is taken to be constant for simplicity.} Variation of the action $S_W$ gives,
\begin{align}
    & \hat{R}_{AB} - \frac{\hat{R}}{2}\hat{g}_{AB} = 0,
\end{align}
as expected. The Neumann boundary condition on $Q_i$ is given as,
\begin{align}
    \hat{K}^{(i)}_{AB} = (\hat{K}^{(i)}-T^{(i)})\hat{h}^{(i)}_{AB}. \label{Nbc}
\end{align}
For the bulk metric \eqref{Bondi_in_hyper}, the outgoing normal at $Q^{(i)}$ is simply,
\begin{align*}
    \hat{n}^{(i)}_A = (-1,0,0),
\end{align*}
and the extrinsic curvature of the surface can be calculated to be,
\begin{align*}
    \hat{K}^{(i)} = \frac{2}{\tau_i}.
\end{align*}
Thus from Israel Junction condition, the tension on the slice is given by:
\begin{align}
    T^{(i)} = \frac{1}{\tau_i}.
\end{align}
The coordinates in $W$ are denoted as $\hat{x}^A\equiv\{\tau,x^\mu\}$, where $x^{\mu}\equiv\{\rho,\theta\}$ are the coordinates on the $AdS_2$ slices. Now we want to perform a consistent dimensional reduction to write down a low energy effective action on the AdS$_2$ slices. In principle, this requires us to vary all components of the bulk metric around the vacuum solution. But since we are interested only in the massless modes in the 2D effective theory, we can choose the following ansatz:
\begin{align}
    ds^2 = -d\tau^2 + \tau^2 g_{\mu\nu}(x)dx^\mu dx^\nu. \label{red_ansatz}
\end{align}
It was shown by \cite{randall_alternative_1999,randall_large_1999,karch_locally_2001} that the modes coming from the variation of cross-terms $g_{\tau\mu}$ correspond to massive modes in the effective theory and hence can be neglected. The massless mode coming from the fluctuation of the $g_{\tau\tau}$ component can be interpreted as the fluctuation of the location of the branes. We will consider this in the next section. \\
~~\\
For this ansatz, we have the following expressions,
\begin{align*}
    & \hat{R}(\hat{g}) = \frac{1}{\tau^2}(R(g)+2), \quad \sqrt{\hat{h}^{(i)}} = \tau_i^2 \sqrt{g}, \quad \hat{K}^{(i)} = \frac{2}{\tau_i}.
\end{align*}
Using these relations in \eqref{actW} and then performing the $\tau$ integral, we obtain the low energy effective theory as follows,
\begin{align}\label{eff-rigid}
    S_{\text{eff}} = \frac{\tau_2 - \tau_1}{16\pi G_N} \int d^2x \sqrt{g} R.
\end{align}
This is nothing but the Einstein-Hilbert action in two dimensions, with an effective 2D Newton constant given as,
\begin{align}
    G_2 \equiv \frac{G_N}{\tau_2 - \tau_1}.
\end{align}
The 2D effective action \eqref{eff-rigid} we got is proportional to the Euler character of the manifold, which is a topological invariant. It has no dynamics. This can also be understood from the fact that the 2D Einstein tensor is trivially zero by construction\footnote{This condition enforces any cosmological constant term to be zero in pure 2D gravity.}.  Although in 3D, the asymptotically flat spacetimes are solutions to the Einstein equation which in turn fixes the metric on the 2D slices to be asymptotically $AdS_2$, we cannot get this condition by simply varying the 2D pure gravity action. Thus as long as superrotation symmetry is there (which is the case for rigid branes), we get no non-trivial dynamics.
\vspace{10pt}\\
To  break this superrotation symmetry, we consider small fluctuations in the location of the branes. As we see below, the fluctuations non-minimally couple to gravity in the effective theory. We essentially use a similar technique to \cite{geng_aspects_2022, geng_jackiw-teitelboim_2022}, where brane fluctuations were considered in AdS$_3$ wedge holography. The authors show that these fluctuations are important in the understanding of entanglement entropy. Brane fluctuations in AdS$_3$ and its effects in entanglement entropy were also studied in \cite{Deng:2022yll}.

\begin{figure}[h] \label{fluctuations}
\centering
\includegraphics[width=1\textwidth]{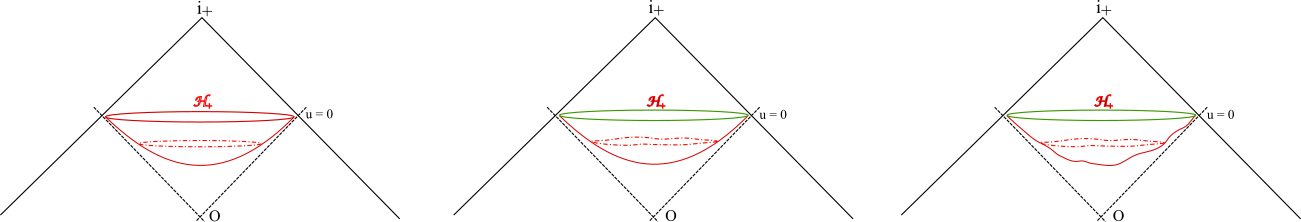}
\caption{The EoW brane and its fluctuations. (a) is a fixed AdS$_2$ brane. (b) shows a rigid brane with spatial fluctuations. (c) shows fluctuating brane configuration. The green circle denotes the fixed metric at the boundary due to Dirichlet boundary condition.}
\end{figure}

\subsection{Reduction to fluctuating AdS$_2$ slices}
We now consider small fluctuations in the location of the $AdS_2$ hypersurfaces, and choose the boundary conditions such that the tension of the branes are held fixed\footnote{This fixes the cosmological constant on the $AdS_2$ slices.}. Again, in low energy description, we consider this additional mode to be independent of $\tau$. The location of the boundary $Q_i$ of $W$ is now given by,
\begin{align}\label{adding-fluc}
    \tau = \tau_i (1+\phi_i(x)).
\end{align}
$\phi_i(x)$ are dimensionless fields that we consider to be very small and we will work up to quadratic order in these fields. The bulk computation remains the same, where only the $\tau$-integration limits change. On the branes we have,
\begin{align*}
    & \hat{n}^{(i)}_A = \left(1+\frac{1}{2}(\nabla\phi_i)^2\right)(-1,\tau_i\nabla_\mu\phi_i), \\
    & \hat{K}^{(i)} = \frac{2}{\tau_i} + \frac{1}{\tau_i}(\nabla^2\phi_i - 2\phi_i) + \frac{2}{\tau_i}(\phi_i^2-\phi_i\nabla^2\phi_i), \\
    & \sqrt{\tilde{h}^{(i)}} = \tau_i^2 \sqrt{g} \left(1 + 2\phi_i + \phi_i^2 - \frac{1}{2}(\nabla\phi_i)^2\right).
\end{align*}

We use these expressions in \eqref{actW} and then perform the $\tau$ integral, to get the effective action up to quadratic order in $\phi$,
\begin{align}
    S_{\text{eff}} =& \frac{\tau_2 - \tau_1}{16\pi G_N} \int d^2x \sqrt{g} R + \frac{1}{16\pi G_N} \int d^2x \sqrt{g}[(\tau_2\phi_2 - \tau_1\phi_1)(R+2) + 2(\tau_1\nabla^2\phi_1 - \tau_2\nabla^2\phi_2)] \nonumber \\
    & + \frac{1}{16\pi G_N} \int d^2x \sqrt{g}[\tau_2(\nabla\phi_2)^2 + 2\tau_2\phi_2^2 - \tau_1(\nabla\phi_1)^2 - 2\tau_1\phi_1^2].
\end{align}
We already see a dilaton gravity theory as the low energy effective theory for the wedge.
\subsubsection{Effective action for the whole hyperbolic patch} \label{Weyl_bulk}
Finally we take the limits $\tau_1 \rightarrow 0$ and $\tau_2 = \tau_\infty$ for large $\tau_\infty$ such that we cover the full hyperbolic patch. Writing $\phi_2\equiv\phi$, the effective action is given by,
\begin{align}
    S_{\text{eff}} =& \frac{\tau_\infty}{16\pi G_N} \int d^2x \sqrt{g}[ R + \phi(R+2) - 2\nabla^2\phi + (\nabla\phi)^2 + 2\phi^2].
\end{align}
We further perform a Weyl transformation of the metric $g\rightarrow \text{e}^{-\phi}g$, such that the kinetic and mass term of $\phi$ gets absorbed in the Ricci scalar,
\begin{align}
    S_{\text{eff}} =& \frac{\tau_\infty}{16\pi G_N} \int d^2x \sqrt{g}[ R + \phi(R+2) + \nabla_\mu(\phi\nabla^\mu\phi - \nabla^\mu\phi)]. \label{Seff_weyl}
\end{align}
This is the bulk part of the JT action up to a total derivative term. In this action, we have dynamical gravity. The dilaton and metric equations of motion are,
\begin{align}
    & R+2 = 0, \\
    & \nabla_\mu\nabla_\nu\phi - g_{\mu\nu}\nabla^2\phi + g_{\mu\nu}\phi = 0. \label{phi-eom}
\end{align}
Thus we see that the dilaton EOM fixes the scalar curvature of the 2D slices to -2. The solutions are asymptotically $AdS_2$ spacetimes. All these solutions are exact zero modes for a constant dilaton profile. But the dilaton profile can in principle be non-trivial at the bulk which makes these modes become slightly nondegenerate (up to the $SL(2,R)$ isometry of global $AdS_2$) in presence of appropriate boundary term. In the 3D picture, it implies that a wedge region with fluctuating boundary breaks the degeneracy between 3D asymptotically flat spacetimes.

\subsubsection{Fall-off condition for the dilaton}
As we just saw, the advent of the dilaton in the effective theory is due to fluctuations of the branes that bound the wedge region. The asymptotic geometry of these branes is fixed from asymptotic flatness condition of (2+1)D. But we still need to impose a non-trivial boundary condition for the dilaton. As it has been extensively studied, the general fall-off condition for the dilaton, consistent with the EOMs in 2D, is given by:
\begin{align}
    \phi(\tilde{\rho} \rightarrow \infty,\theta) = P(\theta)\text{e}^{\tilde{\rho}} + \mathcal{O}(\text{e}^{-\tilde{\rho}}).
\end{align}
It has been discussed in \cite{maldacena_conformal_2016} that through a redefinition of the boundary time coordinate, the $\theta$ dependence from the boundary value of the dilaton can be stripped off. Therefore at leading order, we will consider the following asymptotic behavior of the dilaton,
\begin{align}
    \phi  \xrightarrow{\tilde{\rho}\rightarrow\infty}  \phi_r \text{e}^{\tilde{\rho}}, \label{dil_falloff}
\end{align}
where, $\phi_r$ is a constant. This boundary condition is important for getting a non-trivial dynamics at the boundary.\\

\subsection{Fluctuating dS$_2$ patch: Analytic continuation}
The effective action on the Rindler wedge, which is bounded by two Lorentzian dS$_2$ branes can be calculated quite similarly. The fields on dS$_2$ brane are analytic continuations from the EAdS slice. Along with the continuation of coordinates \eqref{hyper2dS}, we now have the following identifications:
\begin{align}
    L_{AdS} \rightarrow -iL_{dS}, \qquad \phi \rightarrow \psi,
\end{align}
where $L_{(A)dS}$ is the radius of the (A)dS brane and $\psi$ is the scalar mode for fluctuations in the location of the dS$_2$ slice. Thus the effective action on the dS$_2$ brane can be written as,
\begin{align}\label{ds-bulk-action}
    S_{eff} \sim \frac{\xi_{\infty}}{16\pi G_N} \int d^2x \sqrt{-g} [R + \psi (R-2)].
\end{align}
Here once again we have chosen the limits of the wedge region to cover the whole Rindler patch. Similar to \eqref{Seff_weyl}, this action also has an inconsequential total derivative term with it.\\
The analytic continuation also helps fix the boundary behaviour of the metric and dilaton field. Although, dS$_2$ have two conformal boundaries, in \cite{Strominger:2001pn} it was shown that specifying the boundary conditions in either one is sufficient since they are anti-podally identified. Since we are interested in the behaviour near future null infinity in 3D, we will study the boundary behaviour of dS$_2$ fields near future conformal boundary. The conditions are,
\begin{align}
    ds^2 &\xrightarrow{\tilde{\eta} \rightarrow \infty} -d\tilde{\eta}^2 + (\frac{1}{4}e^{2\tilde{\eta}} +\frac{M}{2}+...)d\theta^2, \\
    \psi &\xrightarrow{\tilde{\eta} \rightarrow \infty} i\phi_r e^{\tilde{\eta}}.
\end{align}
The coordinate $\tilde{\eta}$ is given as $\tilde{\eta} \equiv \eta + \ln{\sqrt{M}}$, where $\eta$ is defined through \eqref{hyper2dS}. As we see below these boundary behavior crucially fixes the effective dynamics.

\section{Boundary Schwarzian theory} \label{bdy-theory}
Now that we have understood the effective action on the hyperbolic/dS$_2$ slices, we begin to formulate 1D dual theory as advertised earlier. Both the 3D pure gravity theory and 2D JT gravity are topological theories, hence their dynamics crucially depends on boundary degrees of freedom. In this section, we carefully calculate the boundary action on $u=0$ slice of null infinity.

\subsection{Null Boundary term for 3D gravity}
In general for Einstein-Hilbert action, we need to add a boundary term for a well defined variational principle. In a region with spacelike or timelike boundary, one usually adds a Gibbons-Hawking-York term but in our case the boundary is null. So the usual prescription fails. \cite{parattu_boundary_2016} has a relatively recent proposal for a counter term to be added at null boundaries. We briefly sketch the construction without proof and use it to calculate the necessary boundary term for us.

\subsubsection{General Construction}
For a codimension-1 null hypersurface defined by $\varphi(x) = 0$, the null vector ${l_A\equiv\lambda(x)\partial_A\varphi}$ is normal to the surface. Unlike non-null hypersurfaces, there is no notion of a unit normal since its norm is zero. The usual projector, orthogonal to the normal direction, also does not work since the induced metric on the codimension-1 hypersurface is degenerate.

The vector $l^A$ is also tangent to the null hypersurface and thus we can define integral curves along the surface corresponding to $l^A$ given by $\frac{dx^A}{d\alpha} = l^A$. Along these curves, $\theta$ is constant. Thus $\alpha$ parameterises the null geodesics and $\theta$ represents the coordinates on the spatial slice of the null hypersurface. The hypersurface can be covered by coordinates $\{\alpha,\theta\}$. Depending on the parameterisation $\alpha$, the null geodesics described above satisfy $l^A\nabla_A l_B = \kappa l_B$, where $\kappa$ is the inaffinity.

It is possible to define a nondegenerate induced metric on this spatial slice, which in case of 3D is just one dimensional. To do so, a linearly independent basis should be constructed, given by $\{l^A,k^A,E^A_\theta\}$ such that,
\begin{align}
    l_A l^A = k_A k^A = 0; \quad l_A k^A = -1; \quad E^A_\theta = \frac{\partial x^A}{\partial\theta}; \quad l_A E^A_\theta = k_A E^A_\theta = 0.
\end{align}
Here $k^A$ is an auxiliary null vector. Now the induced metric on the spatial slice can be defined as,
\begin{align}\label{induced-spatial}
    q_{AB} = g_{AB} + l_A k_B + k_A l_B.
\end{align}
$q^A_B$ acts as a projector on the spatial slice, orthogonal to both $l^A$ and $k^A$. Equipped with this, \cite{parattu_boundary_2016} constructed a boundary term that is suitable to make the variation of the Einstein-Hilbert action well-defined on a manifold with null boundary. We are imposing Dirichlet boundary condition on the metric. Then the boundary action is given by,
\begin{align}
     S_{\text{bdy}}=\frac{1}{8\pi G_N}\int d\alpha d\theta\sqrt{q}(\Theta + \kappa).
\end{align}
Here $\Theta = g^{CD}q^A_C q^B_D \nabla_C l_D$ is called the \textit{expansion scalar} and $q$ is the determinant of the metric $q_{AB}$, projected on the spatial slice.

\subsubsection{Null boundary term from 3D asymptotically flat spaces}
Now we want to compute the contribution coming from the null boundary term in the wedge region $W$, bounded by hyperbolic slices. The location of the EoW branes dictate the integration range for the null geodesic parameter. We go to a double null coordinates $(u,v,\theta)$ from $(u,r,\theta)$ using $r = \frac{1}{2}(v - M u)$ such that the metric \eqref{3D_Bondi} for $J = 0$ takes the form,
\begin{align}
    ds^2 = -dudv  +\frac{1}{4}(v-Mu)^2 d\theta^2.
\end{align}
The null boundary $\mathscr{I}^+$ is located at $v = v_\infty$ such that $v_\infty \rightarrow \infty$. Normal to this boundary
\begin{align}
    l_A = \lambda \partial_A(v-v_\infty) = \lambda(0,1,0), \quad l^A = (-2\lambda,0,0).
\end{align}
We have put an arbitrary constant $\lambda$ in front of the normal since the normalization cannot be fixed. The coordinates adapted to the null geodesics on $\mathscr{I}^+$ is given as $(u,\theta)$ such that,
\begin{align}
    \frac{d x^A}{du} \propto l^A.
\end{align}
The null geodesics are affinely parametrized w.r.t $u$ since $l^A\nabla_A l_B = 0$ i.e. the inaffinity parameter $\kappa = 0$.
The auxilliary null vector for this surface is then given by,
\begin{align}
    & k^A k_A = 0, \quad k^A l_A = -1, \quad k_A \frac{\partial x^A}{\partial\theta} = 0, \nonumber \\
    \implies & k_A = \left(\frac{1}{2\lambda},0,0\right), \quad k^A = \left(0,-\frac{1}{\lambda},0\right).
\end{align}
Now we have the necessary quantities to calculate the induced metric on the spatial slice. Using \eqref{induced-spatial} we get,
\begin{align}
    & q_{AB} = g_{AB} + l_A k_B + k_A l_B = \text{diag}\left(0,0,\frac{1}{4}(v-Mu)^2\right); \quad \sqrt{q} = \frac{1}{2}(v - Mu).
\end{align}
We can also calculate the expansion scalar to be,
\begin{align}
\Theta = \frac{2\lambda M}{v - Mu}.
\end{align}
As discussed above, the boundary term to be added to \eqref{Einsbulk} for a well-defined variational principle with Dirichlet boundary condition is given by, 
\begin{align}
    S_{\text{bdy}}=\frac{1}{8\pi G_N}\int du d\theta\sqrt{q}(\Theta + \kappa) = \frac{\lambda}{8\pi G_N} \int d\theta M \int_{u_1}^{u_2} du = \frac{\lambda}{8\pi G_N} \int d\theta M (u_2 - u_1), \label{null_bdy}
\end{align}
where the limits of the $u$-integration are determined by the cut-off on the fluctuating EoW branes. We introduce this cutoff via (large) constant $v$ surfaces. In hyperbolic coordinates these are given by,
\begin{align}
    v = Mu+2r = \tau\text{e}^{\tilde{\rho}} = v_{\infty},
\end{align}
where the boundary $\mathscr{I}^+$ is given by $v_{\infty} \rightarrow \infty$.
This cutoff inherits a cutoff along the radial direction for the fluctuating $AdS_2$ branes. For the brane given by $\tau = \tau_i(1+\phi_i)$ the radial cutoff is at $\tilde{\rho}_i$ such that,
\begin{align}
   \tau_i(1+\phi_i) \text{e}^{\tilde{\rho}_i} = v_\infty. \label{2D_cutoff}
\end{align}
Then from the coordinate transformation relations we have,
\begin{align}
    & u_i = \tau_i(1+\phi_i)\text{e}^{-\tilde{\rho}_i},
\end{align}
as the limits of the above integral. The null boundary term \eqref{null_bdy} becomes,
\begin{align}
    S_{\text{bdy}}=\frac{\lambda}{8\pi G_N}\int d\theta M \left( \tau_2\text{e}^{-\tilde{\rho}_2} - \tau_1 \text{e}^{-\tilde{\rho}_1} + \tau_2\phi_2\text{e}^{-\tilde{\rho}_2} - \tau_1\phi_1 \text{e}^{-\tilde{\rho}_1} \right).
\end{align}
We take the limits $\tau_2 = \tau_\infty$ and $\tau_1 \rightarrow 0$ and we rewrite $\phi_2 = \phi$ and $\tilde{\rho}_2 = \tilde{\rho}_\infty$. For the boundary behavior of the dilaton \eqref{dil_falloff},
\begin{align}
    \phi = \phi_r \text{e}^{\tilde{\rho}_\infty},
\end{align}
the null boundary term \eqref{null_bdy} is given as under the limit $\tilde{\rho}_\infty \rightarrow \infty$,
\begin{align}
    S_{\text{bdy}}=\frac{\lambda}{8\pi G_N}\tau_\infty\int d\theta \phi_r M.
\end{align}
Now we show that this term exactly matches with the boundary term from 2D effective theory.

\subsection{GHY term for 2D (A)dS$_2$ slices}
From \eqref{Bondi_in_hyper} we have that on a constant $\tau = \tau_i$ slice, the asymptotically $AdS_2$ metric has the following form,
\begin{align}
    ds^2 = d\tilde{\rho}^2 + \left(\frac{1}{4}\text{e}^{2\tilde{\rho}} -\frac{M}{2} + \mathcal{O}\left(\text{e}^{-2\tilde{\rho}}\right) \right) d\theta^2. \label{aads2}
\end{align}
The boundary of this asymptotically $AdS_2$ spacetime is given by the intersection of the constant $\tau$ slice and the $v=v_\infty$ surface. Therefore it is located at a constant radial distance $\tilde{\rho} = \tilde{\rho}_i$ given by \eqref{2D_cutoff}  with Dirichlet boundary condition on the metric. For the metric \eqref{aads2}, we can construct the unit normal $n_\mu$, induced metric $\gamma_{\mu\nu}$, and extrinsic curvature $K$ as,
\begin{align*}
    & n_\mu = (1,0), \\
    & \sqrt{\gamma} = \left(\frac{1}{4}\text{e}^{2\tilde{\rho}} -\frac{M}{2}\right)^{1/2}, \\
    & K = \frac{\text{e}^{2\tilde{\rho}}}{\text{e}^{2\tilde{\rho}} - 2M}, \\
    & \sqrt{\gamma} K = \frac{1}{2}\text{e}^{\tilde{\rho}} + \frac{M}{2}\text{e}^{-\tilde{\rho}} + \mathcal{O}\left(\text{e}^{-2\tilde{\rho}}\right).
\end{align*}
Using the falloff \eqref{dil_falloff} of the dilaton, we have
\begin{align}
    \int dx \sqrt{\gamma} \phi(K - 1) = \frac{\phi_r}{2} \int d\theta M(\theta).
\end{align}
Using this condition, the null boundary term of 3D gravity in the hyperbolic region with asymptotic flat boundary conditions, reduces to the GHY boundary term required for the 2D action \eqref{Seff_weyl},
\begin{align}
    I_{\text{Milne}}=\frac{\tau_\infty}{8\pi G_N}\int \sqrt{\gamma} \phi(K - 1) = \frac{\tau_\infty\phi_r}{16\pi G_N} \int d\theta M(\theta). \label{1D_bdy}
\end{align}
We have made the choice $\lambda = \frac{1}{2}$, which was arbitrary.
\vspace{10pt}\\
The GHY term required for the effective action \eqref{ds-bulk-action} in Rindler region, can be similarly computed from the null boundary term,
\begin{align} \label{bdy-ds}
    I_{\text{Rindler}}= -\frac{\xi_\infty}{8\pi G_N}\int \sqrt{\gamma} \psi(K - 1) = -\frac{i\xi_\infty\phi_r}{16\pi G_N} \int d\theta M(\theta).
\end{align}
Using $\xi_\infty = i\tau_\infty$, we find that this contribution is exactly equal to the boundary term \eqref{1D_bdy} and they add up at $u=0$.

\subsection{1D effective action: Schwarzian}
The effective theory that describes the dynamics of 3D asymptotically flat spacetimes (which are related to the flat spacetime through superrotations) in the hyperbolic region, is given by \eqref{Seff_weyl} along with the boundary contribution \eqref{1D_bdy},
\begin{align}
    & S_{\text{eff}} = \frac{1}{16\pi G_2} \int d^2x \sqrt{g} R + S_{\text{JT}}. 
\end{align}
$G_2 = \frac{G_N}{\tau_\infty}$ is the effective two-dimensional Newton constant. In Kaluza-Klein reduction also, the effective lower dimensional coupling constant is given by a combination of the volume of the compact manifold and the higher dimensional coupling. In our non-compact reduction, we have regulated the "infinite box" to a finite size through the introduction of a large cutoff $\tau_\infty$. Similar to Kaluza-Klein reduction, the effective coupling depends on this cutoff i.e. the size of the "finite box".

The first term in the action is a constant and the second term in the JT action given as,
\begin{align}
        & S_{\text{JT}} = \frac{1}{16\pi G_2} \left[ \int d^2x \sqrt{g}  \phi (R+2) + 2 \int dx \sqrt{\gamma} \phi(K-1) \right]. \label{JT}
\end{align}
We have dropped off the total derivative piece from \eqref{Seff_weyl} since its variation is zero in the boundary with our boundary conditions. Hence this term does not affect the dynamics of the system. 

In the JT action \eqref{JT}, the dilaton $\phi$ is a Lagrange multiplier, hence it can be simply integrated out by plugging the dilaton equation of motion into the action, which sets ${R+2 = 0}$ \cite{maldacena_conformal_2016}. This corresponds to asymptotically $AdS_2$ geometries and gives an effective one-dimensional Schwarzian theory coming from the boundary term \eqref{1D_bdy}. As expected, this is consistent with the asymptotic $AdS_2$ boundary conditions that we have obtained from the superrotated 3D spacetimes.

The boundary theory describes the dynamics of the boundary graviton in 2D, which we have identified to be coming from the superrotation mode of 3D gravity. We have already seen that the function $M(\theta)$ transforms as an infinitesimal Schwarzian derivative under the action of 3D superrotation generators. From the 2D perspective \cite{grumiller_menagerie_2017}, it was shown that the $\mathcal{O}(1)$ correction appearing in \eqref{aads2}, transforms as an infinitesimal Schwarzian derivative under the action of the asymptotic Killing vectors of $AdS_2$.
\vspace{10pt}\\
An exact replica of this calculation would occur for dS$_2$ slices. After integrating out $\psi$ from the effective action \eqref{ds-bulk-action} we get $R-2 =0$ \cite{Cotler:2019nbi,Maldacena:2019cbz}. In this case also, the theory reduces to a similar 1D boundary theory as in \eqref{bdy-ds}.  
Thus the effective action (considering contributions from both Milne and Rindler patches) at the $u=0$ circle that describes the low energy dynamics of 3D asymptotically flat spaces is given by,
\begin{align}
    S_{1D} = I_{\text{Milne}} + I_{\text{Rindler}} = \frac{\phi_r}{8\pi G_2} \int d\theta M(\theta). \label{1D_eff}
\end{align}
Due to time translation symmetry in 3D, this effective theory may live on any spatial slice of $\mathscr{I}^+$. For superrotated spacetimes in 3D, this action can be thought of as the (nearly) Celestial CFT dual of the 3D pure gravity theory. This effective action \eqref{1D_eff} is closely related to the action of superrotation Goldstone modes derived in \cite{carlip_dynamics_2017} through a different prescription.
\vspace{10pt}\\
An exactly similar effective action can be written down at $v=0$ surface of past null infinity $\mathscr{I}_-$. There the Schwarzian theory would be described in terms of $\tilde{M}$, the superrotation modes at $\mathscr{I}_-$. But this is not an independent theory as $\tilde{M}$ is related to $M(\theta)$ via antipodal matching condition. Hence the 1D theory can be described either at $\mathscr{I}_+$ or $\mathscr{I}_-$. In the dimensionally reduced picture this implies that the hyperbolic slices on $\mathcal{H}^+$ and $\mathcal{H}^-$ have boundary conditions that are antipodally matched. For dS$_2$ slices, antipodal matching of 3D implies the boundary conditions on future and past conformal boundary to be identified (as expected in \cite{Strominger:2001pn}).
\section{Conclusions and Outlook} \label{concl}
In this paper we study the low energy effective dynamics of pure gravity in asymptotically flat spacetime. We start by reducing the 3D theory on a wedge region bounded by two asymptotically (A)dS$_2$ slices in Milne and Rindler patches separately. Finally we take the limit when the wedge regions cover the full spacetime. The localised action at these slices turn out to be JT gravity when fluctuations of the EoW branes are taken into account. Using the dual boundary description of JT gravity, we find that the co-dimension two holographic dual is a Schwarzian theory that lives at a spatial slice of $\mathscr{I}^+$. The superrotation mode in 3D acts as the Schwarzian in this action. In this process, we identify that the Virasoro subalgebra of BMS$_3$, generated by the superrotations, maps to the asymptotic symmetry algebra of (A)dS$_2$. Evidently, this is nothing but the local conformal algebra (or diffeomorphisms) of the 1D boundary. This is an explicit construction of a Celestial (nearly) CFT in low energy limit.
\vspace{10pt}\\
The $\phi$ mode we have introduced to break superrotation symmetry needs to be understood more. It clearly has non-trivial boundary value and hence behaves as a large diffeomorphism. The equation of motion \eqref{phi-eom} suggests that $\phi$ can be interpreted as a large diffeomorphism in harmonic gauge. We hope to explore this further in a future work.
\vspace{10pt}\\
There are several directions to explore from here. Firstly, we only considered superrotated spacetimes in 3D as those solutions respect the hyperbolic foliation crucial for the construction. The effect of 3D supertranslations on this effective theory needs to be understood and we want to explore this further. Understanding the 1D dual theory in presence of fermionic symmetries of (2+1)D supergravity theories \cite{Barnich:2015sca, Fuentealba:2017fck, Banerjee:2019lrv, Lodato:2016alv, Banerjee:2017gzj} would be interesting. The 1D dual picture we have, also must relate to the 2D Wess-Zumino-Witten/Liouville dual description of 3D gravity (See \cite{Barnich:2012rz} for pure gravity and \cite{Barnich:2015sca, Banerjee:2019lrv, Banerjee:2021uxl} for supergravity). The relation between Lioville theories and Schwarzian are studied in \cite{Mertens:2017mtv, Mertens:2018fds}. Also, the explicit relations between correlators in 3D flat space and the ones coming from the Schwarzian action needs to be understood further. 
\vspace{10pt}\\
In \cite{Barnich:2015mui} the one loop partition function for 3D flat space was calculated, shown to be related to the characters of the BMS$_3$ group. It would be interesting to understand this partition function from the 1D effective theory point of view. It has been shown that JT gravity is dual to a random matrix model \cite{Saad:2019lba}. It is worth exploring if there is some correspondence between 3D gravity with matrix models, via its duality with the Schwarzian.  Supertranslations may non-trivially affect this description. Along this line, we would also like to understand whether there is a connection with the construction of BMS$_3$ invariant matrix models \cite{Bhattacharjee:2021zju}. 

\acknowledgments

We are grateful to Nabamita Banerjee, Dileep Jatkar, Alok Laddha and Tadashi Takayanagi for illuminating discussions and useful comments on the draft. MS would like to thank Tabasum Rahnuma for helpful discussions. We would like to thank the people of India for their continuous support towards research in basic sciences.


\bibliography{3DtoJT}

\providecommand{\href}[2]{#2}\begingroup\raggedright\begin{thebibliography}{10}

\bibitem{Strominger:2017zoo}
A.~Strominger, {\it {Lectures on the Infrared Structure of Gravity and Gauge
  Theory}},  \href{http://arxiv.org/abs/1703.05448}{{\tt arXiv:1703.05448}}.

\bibitem{Pasterski:2016qvg}
S.~Pasterski, S.-H. Shao, and A.~Strominger, {\it {Flat Space Amplitudes and
  Conformal Symmetry of the Celestial Sphere}},  {\em Phys. Rev. D} {\bf 96}
  (2017), no.~6 065026, [\href{http://arxiv.org/abs/1701.00049}{{\tt
  arXiv:1701.00049}}].

\bibitem{Laddha:2020kvp}
A.~Laddha, S.~G. Prabhu, S.~Raju, and P.~Shrivastava, {\it {The Holographic
  Nature of Null Infinity}},  {\em SciPost Phys.} {\bf 10} (2021), no.~2 041,
  [\href{http://arxiv.org/abs/2002.02448}{{\tt arXiv:2002.02448}}].

\bibitem{Kapec:2016jld}
D.~Kapec, P.~Mitra, A.-M. Raclariu, and A.~Strominger, {\it {2D Stress Tensor
  for 4D Gravity}},  {\em Phys. Rev. Lett.} {\bf 119} (2017), no.~12 121601,
  [\href{http://arxiv.org/abs/1609.00282}{{\tt arXiv:1609.00282}}].

\bibitem{Barnich:2009se}
G.~Barnich and C.~Troessaert, {\it {Symmetries of asymptotically flat 4
  dimensional spacetimes at null infinity revisited}},  {\em Phys. Rev. Lett.}
  {\bf 105} (2010) 111103, [\href{http://arxiv.org/abs/0909.2617}{{\tt
  arXiv:0909.2617}}].

\bibitem{Pasterski:2017kqt}
S.~Pasterski and S.-H. Shao, {\it {Conformal basis for flat space amplitudes}},
   {\em Phys. Rev. D} {\bf 96} (2017), no.~6 065022,
  [\href{http://arxiv.org/abs/1705.01027}{{\tt arXiv:1705.01027}}].

\bibitem{Barnich:2010ojg}
G.~Barnich and C.~Troessaert, {\it {Supertranslations call for
  superrotations}},  {\em PoS} {\bf CNCFG2010} (2010) 010,
  [\href{http://arxiv.org/abs/1102.4632}{{\tt arXiv:1102.4632}}].

\bibitem{Campiglia:2014yka}
M.~Campiglia and A.~Laddha, {\it {Asymptotic symmetries and subleading soft
  graviton theorem}},  {\em Phys. Rev. D} {\bf 90} (2014), no.~12 124028,
  [\href{http://arxiv.org/abs/1408.2228}{{\tt arXiv:1408.2228}}].

\bibitem{Kapec:2014opa}
D.~Kapec, V.~Lysov, S.~Pasterski, and A.~Strominger, {\it {Semiclassical
  Virasoro symmetry of the quantum gravity $ \mathcal{S}$-matrix}},  {\em JHEP}
  {\bf 08} (2014) 058, [\href{http://arxiv.org/abs/1406.3312}{{\tt
  arXiv:1406.3312}}].

\bibitem{Barnich:2010eb}
G.~Barnich and C.~Troessaert, {\it {Aspects of the BMS/CFT correspondence}},
  {\em JHEP} {\bf 05} (2010) 062, [\href{http://arxiv.org/abs/1001.1541}{{\tt
  arXiv:1001.1541}}].

\bibitem{deBoer:2003vf}
J.~de~Boer and S.~N. Solodukhin, {\it {A Holographic reduction of Minkowski
  space-time}},  {\em Nucl. Phys. B} {\bf 665} (2003) 545--593,
  [\href{http://arxiv.org/abs/hep-th/0303006}{{\tt hep-th/0303006}}].

\bibitem{Cheung:2016iub}
C.~Cheung, A.~de~la Fuente, and R.~Sundrum, {\it {4D scattering amplitudes and
  asymptotic symmetries from 2D CFT}},  {\em JHEP} {\bf 01} (2017) 112,
  [\href{http://arxiv.org/abs/1609.00732}{{\tt arXiv:1609.00732}}].

\bibitem{akal_codimension_2020}
I.~Akal, Y.~Kusuki, T.~Takayanagi, and Z.~Wei, {\it {Codimension two holography
  for wedges}},  {\em Phys. Rev. D} {\bf 102} (2020), no.~12 126007,
  [\href{http://arxiv.org/abs/2007.06800}{{\tt arXiv:2007.06800}}].

\bibitem{Jackiw:1984je}
R.~Jackiw, {\it {Lower Dimensional Gravity}},  {\em Nucl. Phys. B} {\bf 252}
  (1985) 343--356.

\bibitem{Teitelboim:1983ux}
C.~Teitelboim, {\it {Gravitation and Hamiltonian Structure in Two Space-Time
  Dimensions}},  {\em Phys. Lett. B} {\bf 126} (1983) 41--45.

\bibitem{geng_jackiw-teitelboim_2022}
H.~Geng, A.~Karch, C.~Perez-Pardavila, S.~Raju, L.~Randall, M.~Riojas, and
  S.~Shashi, {\it {Jackiw-Teitelboim Gravity from the Karch-Randall
  Braneworld}},  \href{http://arxiv.org/abs/2206.04695}{{\tt
  arXiv:2206.04695}}.

\bibitem{geng_aspects_2022}
H.~Geng, {\it {Aspects of AdS$_{2}$ quantum gravity and the Karch-Randall
  braneworld}},  {\em JHEP} {\bf 09} (2022) 024,
  [\href{http://arxiv.org/abs/2206.11277}{{\tt arXiv:2206.11277}}].

\bibitem{Deng:2022yll}
F.~Deng, Y.-S. An, and Y.~Zhou, {\it {JT Gravity from Partial Reduction and
  Defect Extremal Surface}},  \href{http://arxiv.org/abs/2206.09609}{{\tt
  arXiv:2206.09609}}.

\bibitem{Maldacena:1998uz}
J.~M. Maldacena, J.~Michelson, and A.~Strominger, {\it {Anti-de Sitter
  fragmentation}},  {\em JHEP} {\bf 02} (1999) 011,
  [\href{http://arxiv.org/abs/hep-th/9812073}{{\tt hep-th/9812073}}].

\bibitem{maldacena_conformal_2016}
J.~Maldacena, D.~Stanford, and Z.~Yang, {\it {Conformal symmetry and its
  breaking in two dimensional Nearly Anti-de-Sitter space}},  {\em PTEP} {\bf
  2016} (2016), no.~12 12C104, [\href{http://arxiv.org/abs/1606.01857}{{\tt
  arXiv:1606.01857}}].

\bibitem{carlip_dynamics_2017}
S.~Carlip, {\it {The dynamics of supertranslations and superrotations in 2 + 1
  dimensions}},  {\em Class. Quant. Grav.} {\bf 35} (2018), no.~1 014001,
  [\href{http://arxiv.org/abs/1608.05088}{{\tt arXiv:1608.05088}}].

\bibitem{Bondi:1962px}
H.~Bondi, M.~G.~J. van~der Burg, and A.~W.~K. Metzner, {\it {Gravitational
  waves in general relativity. 7. Waves from axisymmetric isolated systems}},
  {\em Proc. Roy. Soc. Lond. A} {\bf 269} (1962) 21--52.

\bibitem{Sachs:1962wk}
R.~K. Sachs, {\it {Gravitational waves in general relativity. 8. Waves in
  asymptotically flat space-times}},  {\em Proc. Roy. Soc. Lond. A} {\bf 270}
  (1962) 103--126.

\bibitem{Campiglia:2015yka}
M.~Campiglia and A.~Laddha, {\it {New symmetries for the Gravitational
  S-matrix}},  {\em JHEP} {\bf 04} (2015) 076,
  [\href{http://arxiv.org/abs/1502.02318}{{\tt arXiv:1502.02318}}].

\bibitem{Strominger:2013jfa}
A.~Strominger, {\it {On BMS Invariance of Gravitational Scattering}},  {\em
  JHEP} {\bf 07} (2014) 152, [\href{http://arxiv.org/abs/1312.2229}{{\tt
  arXiv:1312.2229}}].

\bibitem{barnich_dual_2013}
G.~Barnich and H.~A. Gonzalez, {\it {Dual dynamics of three dimensional
  asymptotically flat Einstein gravity at null infinity}},  {\em JHEP} {\bf 05}
  (2013) 016, [\href{http://arxiv.org/abs/1303.1075}{{\tt arXiv:1303.1075}}].

\bibitem{grumiller_menagerie_2017}
D.~Grumiller, R.~McNees, J.~Salzer, C.~Valc\'arcel, and D.~Vassilevich, {\it
  {Menagerie of AdS$_{2}$ boundary conditions}},  {\em JHEP} {\bf 10} (2017)
  203, [\href{http://arxiv.org/abs/1708.08471}{{\tt arXiv:1708.08471}}].

\bibitem{miao_exact_2020}
R.-X. Miao, {\it {An Exact Construction of Codimension two Holography}},  {\em
  JHEP} {\bf 01} (2021) 150, [\href{http://arxiv.org/abs/2009.06263}{{\tt
  arXiv:2009.06263}}].

\bibitem{Geng:2020fxl}
H.~Geng, A.~Karch, C.~Perez-Pardavila, S.~Raju, L.~Randall, M.~Riojas, and
  S.~Shashi, {\it {Information Transfer with a Gravitating Bath}},  {\em
  SciPost Phys.} {\bf 10} (2021), no.~5 103,
  [\href{http://arxiv.org/abs/2012.04671}{{\tt arXiv:2012.04671}}].

\bibitem{Geng:2021iyq}
H.~Geng, S.~L\"ust, R.~K. Mishra, and D.~Wakeham, {\it {Holographic BCFTs and
  Communicating Black Holes}},  {\em jhep} {\bf 08} (2021) 003,
  [\href{http://arxiv.org/abs/2104.07039}{{\tt arXiv:2104.07039}}].

\bibitem{ogawa_wedge_2022}
N.~Ogawa, T.~Takayanagi, T.~Tsuda, and T.~Waki, {\it {Wedge Holography in Flat
  Space and Celestial Holography}},
  \href{http://arxiv.org/abs/2207.06735}{{\tt arXiv:2207.06735}}.

\bibitem{randall_alternative_1999}
L.~Randall and R.~Sundrum, {\it {An Alternative to compactification}},  {\em
  Phys. Rev. Lett.} {\bf 83} (1999) 4690--4693,
  [\href{http://arxiv.org/abs/hep-th/9906064}{{\tt hep-th/9906064}}].

\bibitem{randall_large_1999}
L.~Randall and R.~Sundrum, {\it {A Large mass hierarchy from a small extra
  dimension}},  {\em Phys. Rev. Lett.} {\bf 83} (1999) 3370--3373,
  [\href{http://arxiv.org/abs/hep-ph/9905221}{{\tt hep-ph/9905221}}].

\bibitem{karch_locally_2001}
A.~Karch and L.~Randall, {\it {Locally localized gravity}},  {\em JHEP} {\bf
  05} (2001) 008, [\href{http://arxiv.org/abs/hep-th/0011156}{{\tt
  hep-th/0011156}}].

\bibitem{Strominger:2001pn}
A.~Strominger, {\it {The dS / CFT correspondence}},  {\em JHEP} {\bf 10} (2001)
  034, [\href{http://arxiv.org/abs/hep-th/0106113}{{\tt hep-th/0106113}}].

\bibitem{parattu_boundary_2016}
K.~Parattu, S.~Chakraborty, B.~R. Majhi, and T.~Padmanabhan, {\it {A Boundary
  Term for the Gravitational Action with Null Boundaries}},  {\em Gen. Rel.
  Grav.} {\bf 48} (2016), no.~7 94,
  [\href{http://arxiv.org/abs/1501.01053}{{\tt arXiv:1501.01053}}].

\bibitem{Cotler:2019nbi}
J.~Cotler, K.~Jensen, and A.~Maloney, {\it {Low-dimensional de Sitter quantum
  gravity}},  {\em JHEP} {\bf 06} (2020) 048,
  [\href{http://arxiv.org/abs/1905.03780}{{\tt arXiv:1905.03780}}].

\bibitem{Maldacena:2019cbz}
J.~Maldacena, G.~J. Turiaci, and Z.~Yang, {\it {Two dimensional Nearly de
  Sitter gravity}},  {\em JHEP} {\bf 01} (2021) 139,
  [\href{http://arxiv.org/abs/1904.01911}{{\tt arXiv:1904.01911}}].

\bibitem{Barnich:2015sca}
G.~Barnich, L.~Donnay, J.~Matulich, and R.~Troncoso, {\it {Super-BMS$_{3}$
  invariant boundary theory from three-dimensional flat supergravity}},  {\em
  JHEP} {\bf 01} (2017) 029, [\href{http://arxiv.org/abs/1510.08824}{{\tt
  arXiv:1510.08824}}].

\bibitem{Fuentealba:2017fck}
O.~Fuentealba, J.~Matulich, and R.~Troncoso, {\it {Asymptotic structure of
  $\mathcal{N}=2$ supergravity in 3D: extended super-BMS$_3$ and nonlinear
  energy bounds}},  {\em JHEP} {\bf 09} (2017) 030,
  [\href{http://arxiv.org/abs/1706.07542}{{\tt arXiv:1706.07542}}].

\bibitem{Banerjee:2019lrv}
N.~Banerjee, A.~Bhattacharjee, Neetu, and T.~Neogi, {\it {New $ \mathcal{N} $ =
  2 SuperBMS$_{3}$ algebra and invariant dual theory for 3D supergravity}},
  {\em JHEP} {\bf 11} (2019) 122, [\href{http://arxiv.org/abs/1905.10239}{{\tt
  arXiv:1905.10239}}].

\bibitem{Lodato:2016alv}
I.~Lodato and W.~Merbis, {\it {Super-BMS$_{3}$ algebras from $ \mathcal{N}=2 $
  flat supergravities}},  {\em JHEP} {\bf 11} (2016) 150,
  [\href{http://arxiv.org/abs/1610.07506}{{\tt arXiv:1610.07506}}].

\bibitem{Banerjee:2017gzj}
N.~Banerjee, I.~Lodato, and T.~Neogi, {\it {N=4 Supersymmetric BMS3 algebras
  from asymptotic symmetry analysis}},  {\em Phys. Rev. D} {\bf 96} (2017),
  no.~6 066029, [\href{http://arxiv.org/abs/1706.02922}{{\tt
  arXiv:1706.02922}}].

\bibitem{Barnich:2012rz}
G.~Barnich, A.~Gomberoff, and H.~A. Gonz\'alez, {\it {Three-dimensional
  Bondi-Metzner-Sachs invariant two-dimensional field theories as the flat
  limit of Liouville theory}},  {\em Phys. Rev. D} {\bf 87} (2013), no.~12
  124032, [\href{http://arxiv.org/abs/1210.0731}{{\tt arXiv:1210.0731}}].

\bibitem{Banerjee:2021uxl}
N.~Banerjee, A.~Bhattacharjee, S.~Biswas, and T.~Neogi, {\it {Dual theory for
  maximally $ \mathcal{N} $ extended flat supergravity}},  {\em JHEP} {\bf 05}
  (2022) 179, [\href{http://arxiv.org/abs/2110.05919}{{\tt arXiv:2110.05919}}].

\bibitem{Mertens:2017mtv}
T.~G. Mertens, G.~J. Turiaci, and H.~L. Verlinde, {\it {Solving the Schwarzian
  via the Conformal Bootstrap}},  {\em JHEP} {\bf 08} (2017) 136,
  [\href{http://arxiv.org/abs/1705.08408}{{\tt arXiv:1705.08408}}].

\bibitem{Mertens:2018fds}
T.~G. Mertens, {\it {The Schwarzian theory \textemdash{} origins}},  {\em JHEP}
  {\bf 05} (2018) 036, [\href{http://arxiv.org/abs/1801.09605}{{\tt
  arXiv:1801.09605}}].

\bibitem{Barnich:2015mui}
G.~Barnich, H.~A. Gonzalez, A.~Maloney, and B.~Oblak, {\it {One-loop partition
  function of three-dimensional flat gravity}},  {\em JHEP} {\bf 04} (2015)
  178, [\href{http://arxiv.org/abs/1502.06185}{{\tt arXiv:1502.06185}}].

\bibitem{Saad:2019lba}
P.~Saad, S.~H. Shenker, and D.~Stanford, {\it {JT gravity as a matrix
  integral}},  \href{http://arxiv.org/abs/1903.11115}{{\tt arXiv:1903.11115}}.

\bibitem{Bhattacharjee:2021zju}
A.~Bhattacharjee and Neetu, {\it {Matrix model with 3D BMS constraints}},  {\em
  Phys. Rev. D} {\bf 105} (2022), no.~6 066012,
  [\href{http://arxiv.org/abs/2108.07314}{{\tt arXiv:2108.07314}}].

\end{thebibliography}\endgroup
\bibliographystyle{jhep}
\end{document}